\documentclass[prd,twocolumn,tightenlines,superscriptaddress,nofootinbib,
showpacs]{revtex4}
\usepackage{float}
\usepackage{amssymb,latexsym}
\usepackage{amsmath,amsbsy,bbm}
\usepackage{epsfig,bm}
\usepackage{graphicx,comment}
\usepackage{color}
\begin{document}

\title{Revisiting the Phase Transition of 
Spin-1/2 Heisenberg Model with a Spatially Staggered Anisotropy on the
Square Lattice}

\author{F.-J. Jiang}
\email[]{fjjiang@ntnu.edu.tw}
\affiliation{Department of Physics, National Taiwan Normal University, 88, Sec.4, Ting-Chou Rd., Taipei 116, 
Taiwan}
\vspace{-1cm}
  
\begin{abstract}
Puzzled by the indication of a new critical theory for the
spin-1/2 Heisenberg model with a spatially staggered anisotropy on the
square lattice as suggested in \cite{Wenzel08}, we re-investigate
the phase transition of this model induced by dimerization using first principle Monte
Carlo simulations.
We focus on studying the finite-size scaling of
$\rho_{s1} L$ and $\rho_{s2} L$, where $L$ stands for the spatial box size used in the 
simulations and $\rho_{si}$ with $i \in \{1,2\}$ is the 
spin-stiffness in $i$-direction.
From our Monte Carlo data,
we find that $\rho_{s2} L$ suffers a much less severe correction compared to that of 
$\rho_{s1} L$. Therefore $\rho_{s2} L$ is a better quantity than $\rho_{s1} L$ for
finite-size scaling analysis concerning the limitation of the availability of 
large volumes data in our study. Further, motivated by the so-called cubical regime in
magnon chiral perturbation theory, we additionally perform a finite-size scaling analysis 
on our Monte Carlo data with the assumption that 
the ratio of spatial winding numbers squared is fixed through all
simulations. As a result, the physical shape
of the system remains fixed in our calculations. The validity of
this new idea is confirmed by studying the phase transition driven by spatial anisotropy
for the ladder anisotropic Heisenberg model.  
With this new strategy, even from $\rho_{s1} L$ which receives 
the most serious correction among the observables
considered in this study, we arrive at a value for the critical exponent 
$\nu$ which is consistent with the expected $O(3)$ value by using only up
to $L = 64$ data points.
\end{abstract}

\maketitle

\section{Introduction}
\vskip-0.2cm
Heisenberg-type models have been studied in great detail
during the last twenty years because of their 
phenomenological importance. 
For example, it is believed that the spin-1/2 Heisenberg model on the square 
lattice is the correct model for understanding the undoped precursors of 
high $T_c$ cuprates (undoped antiferromagnets).
Further, due to the availability of efficient Monte Carlo algorithms as well
as the increasing power of computing resources, properties of undoped antiferromagnets
on geometrically non-frustrated lattices have been determined to unprecedented 
accuracy \cite{Sandvik97,Sandvik99,Kim00,Wang05,Jiang08,Alb08,Wenzel09}. 
For instance, using a loop algorithm, the low-energy
parameters of the spin-1/2 Heisenberg model on the square lattice
are calculated very precisely and are in quantitative agreement with the experimental 
results \cite{Wie94}. Despite being well studied, 
several recent numerical investigation of anisotropic Heisenberg models have led to unexpected
results \cite{Wenzel08,Pardini08,Jiang09.1}. In particular, Monte Carlo evidence indicates that the anisotropic
Heisenberg model with staggered arrangement of the antiferromagnetic 
couplings may belong to a new universality class, in contradiction
to the theoretical $O(3)$ universality prediction \cite{Wenzel08}.
For example, while the most accurate Monte Carlo value for the critical exponent
$\nu$ in the $O(3)$ universality class is given by $\nu=0.7112(5)$ \cite{Cam02},
the corresponding $\nu$ determined in \cite{Wenzel08} is shown to be $\nu=0.689(5)$. 
Although subtlety of calculating the critical exponent $\nu$ from performing
finite-size scaling analysis is demonstrated for a similar anisotropic 
Heisenberg model on the honeycomb lattice \cite{Jiang09.2}, the discrepancy between $\nu = 0.689(5)$ 
and $\nu=0.7112(5)$ observed in \cite{Wenzel08,Cam02} remains to be understood.

In order to clarify this issue further, we have simulated the spin-1/2 Heisenberg model with
a spatially staggered anisotropy on the square lattice. Further, we choose to analyze 
the finite-size scaling of the observables $\rho_{s1} L$ and $\rho_{s2} L$,
where $L$ refers to the box size used in the simulations and $\rho_{si}$ with 
$i \in \{1,2\}$ is the spin stiffness in $i$-direction.
The reason for choosing $\rho_{s1} L$ and $\rho_{s2} L$ is twofold. 
First of all, these two observables can be calculated
to a very high accuracy using loop algorithms. Secondly, one can measure 
$\rho_{s1}$ and $\rho_{s2}$ separately. 
In practice, one would
naturally use $\rho_s $ which is the average of $\rho_{s1}$ and $\rho_{s2}$
for the data analysis.
However for the model considered here, 
we find it is useful to analyze both the data of $\rho_{s1}$ and $\rho_{s2}$
because studying $\rho_{s1}$
and $\rho_{s2}$ individually might reveal the impact of anisotropy on the system.
Surprisingly, as we will show later, the observable $\rho_{s2} L$ receives a much
less severe correction than $\rho_{s1} L$ does.
Hence $\rho_{s2} L$ is a better observable than
$\rho_{s1} L$ (or $\rho_{s} L$) for finite-size scaling analysis concerning the limitation of
the availability of large volumes data in this study. 
Further, motivated by the so-called cubical regime in magnon chiral perturbation theory,
we have performed an additional finite-size scaling analysis on $\rho_{s1} L$ with the assumption
that the ratio of spatial winding numbers squared is fixed in all our 
Monte-Carlo simulations. In other word, we keep the physical shape of the
system fixed in the additional analysis of finite-size scaling. The validity of
this new idea is confirmed by studying the phase transition driven by spatial anisotropy
for the ladder anisotropic Heisenberg model, namely the critical point and critical exponent $\nu$ 
for this phase transition we obtain by fixing the ratio of
spatial winding numbers squared are consistent with
the known results in the literature. 
Remarkably, combining the idea of fixing the ratio of spatial winding numbers squared in the 
simulations and finite-size scaling analysis, unlike the unconventional value for
$\nu$ observed in \cite{Wenzel08}, even from $\rho_{s1} L$ which suffers a very serious correction,
we arrive at a value for $\nu$ which is consistent with that of $O(3)$ by using only up to $L=64$ data points.  
   
This paper is organized as follows. In section \ref{model}, the anisotropic
Heisenberg model and the relevant observables studied in this work are briefly described.
Section \ref{results} contains our numerical results.
In particular, the corresponding critical point as well as the critical
exponent $\nu$ are determined by fitting the numerical data to their predicted
critical behavior near the transition. Finally, we conclude our study
in section \ref{discussion}.

\begin{figure}
\begin{center}
\includegraphics[width=0.33\textwidth]{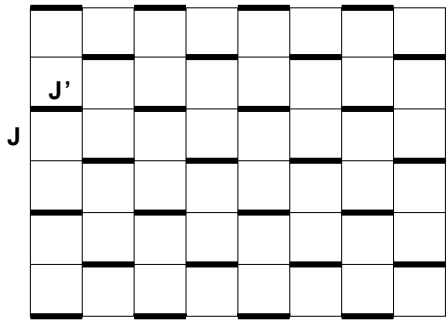}
\end{center}\vskip-0.5cm
\caption{The anisotropic Heisenberg model considered in this study.}
\label{fig0}
\end{figure}

\vskip-0.2cm

\section{Microscopic Model and Corresponding Observables}
\label{model}\vskip-0.2cm 
The Heisenberg
model considered in this study is defined by the Hamilton operator
\begin{eqnarray}
\label{hamilton}
H = \sum_{\langle xy \rangle}J\,\vec S_x \cdot \vec S_{y}
+\sum_{\langle x'y' \rangle}J'\,\vec S_{x'} \cdot \vec S_{y'},
\end{eqnarray}
where $J$ and $J'$ are antiferromagnetic exchange couplings connecting
nearest neighbor spins $\langle  xy \rangle$
and $\langle x'y' \rangle$, respectively. Figure 1 illustrates the Heisenberg
model described by Eq.~(\ref{hamilton}). 
To study the critical behavior of this anisotropic Heisenberg model near 
the transition driven by spatial anisotropy, in particular to determine 
the critical point as well as the critical exponent $\nu$, 
the spin stiffnesses in $1$- and $2$-directions which are defined by\vskip-0.5cm
\begin{eqnarray}
\rho_{si} = \frac{1}{\beta L^2}\langle W^2_{i}\rangle,
\end{eqnarray}
are measured in our simulations.
Here $\beta$ is the inverse temperature and $L$ 
refers to the spatial box size. Further $\langle W^2_{i} \rangle$ 
with $i \in \{1,2\}$ is
the winding number squared in $i$-direction.
By carefully investigating the spatial volumes 
and the $J'/J$ dependence of
$\rho_{s i}L$, one can determine the critical point as well
as the critical exponent $\nu$ with high precision.

\begin{figure}
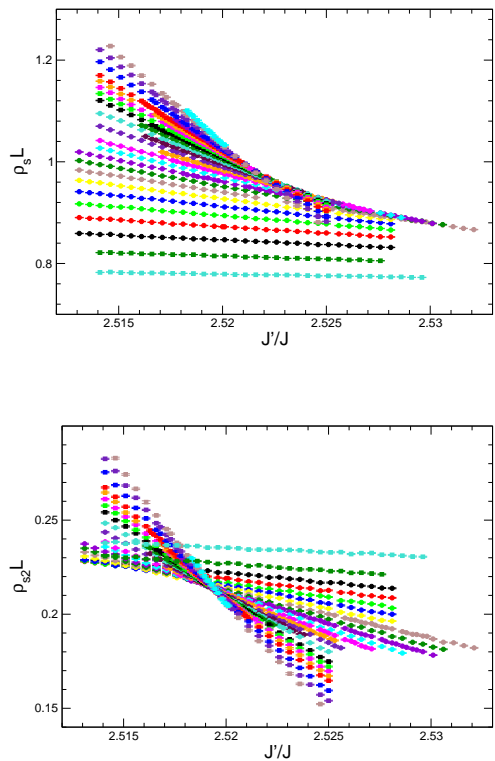

\begin{center}
\vbox{
\includegraphics[width=0.36\textwidth]{rhosL.eps}\vskip1.0cm
\includegraphics[width=0.36\textwidth]{rhos2L2.eps}
}
\end{center}\vskip-0.5cm
\caption{Monte Carlo data of $\rho_{s} L$ (upper panel) and $\rho_{s2} L$ 
(lower panel) as functions of the parameter $J'/J$.}
\label{fig1}
\end{figure}

\begin{figure}
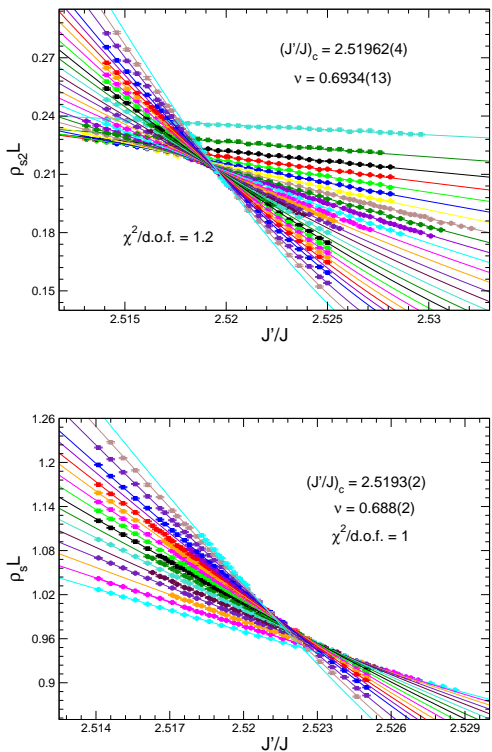

\begin{center}
\vbox{
\includegraphics[width=0.36\textwidth]{rhos2L2fit1.eps}\vskip0.9cm
\includegraphics[width=0.36\textwidth]{rhosLfit1.eps}\vskip0.9cm
}
\end{center}\vskip-0.5cm
\caption{Fits of $\rho_{s2} L$ (upper panel) and $\rho_{s}L $ 
(lower panel) to Eq.~(\ref{FSS}). While the circles and squares on these two panels
are the numerical Monte Carlo data from the simulations, the solid curves 
are obtained by using the results from the fits.}
\label{fig2}
\end{figure}

\vskip-0.2cm

\section{Determination of the Critical Point and the Critical Exponent $\nu$}
\label{results}\vskip-0.2cm
To calculate the relevant critical exponent $\nu$ and to determine the
location of the critical point in the parameter space $J'/J$, one useful technique
is to study the finite-size scaling of certain observables. For example,
if the transition is second order, then near the transition, the observable 
$\rho_{si} L^p$ for $i\in \{1,2\}$ should be described well by the following finite-size scaling 
ansatz
\begin{equation}
\label{FSS}
{\cal O}_{L^p}(t) = ( 1 - b(L^p)^{-\omega} )g_{{\cal O}}(t(L^p)^{1/\nu}), 
\end{equation}
where ${\cal O}_{L^p}$ stands for $\rho_{si}L^p$, $L^p$ is the physical linear length of the system, 
$t = (j_c-j)/j_c$ with $j = (J'/J)$, $b$ is some constant,
$\nu$ is the critical exponent corresponding to the correlation length $\xi$ 
and $\omega$ is the confluent correction exponent.  
Finally $g_{{\cal O}}$ appearing above is a 
smooth function of the variable $t(L^p)^{1/\nu}$. 
In practice, the $L^p$ appearing in Eq. (\ref{FSS}) is conventionally replaced by the box size $L$ used in the simulations when performing finite-size scaling analysis. We will adopt this conventional strategy in the first part of our analysis as well.
From Eq.~(\ref{FSS}), one concludes that 
the curves of different $L$ for ${\cal O}_{L}$, as functions of $J'/J$,
should have the tendency to intersect at critical point $(J'/J)_c$ for large $L$. 
To calculate the critical exponent $\nu$ and the critical point $(J'/J)_c$,
in the following we will apply the finite-size scaling formula,
Eq.~(\ref{FSS}), to both $\rho_{s1} L$ and $\rho_{s2} L$. 
Without losing the generality, in our simulations we have 
fixed $J$ to be $1.0$ and have varied $J'$. Further, the box size used in 
the simulations ranges from $L = 6$ to $L = 64$.
We also use large enough $\beta$ so that the observables studied here 
take their zero-temperature values. Figure \ref{fig1} shows the Monte Carlo data of 
$\rho_{s} L$ and $\rho_{s2} L$ as functions of the parameter $J'/J$. 
The figure clearly indicates the phase 
transition is likely second order since different $L$ curves
for both $\rho_{s} L$ and $\rho_{s2} L$ tend to intersect at a particular point in 
the parameter space $J'/J$. What is the most striking observation from our results
is that the observable $\rho_{s} L$ receives a much severe correction 
than $\rho_{s2} L$ does. This can be understood from the trend of the crossing among these 
curves of different $L$ in figure \ref{fig1}. Therefore one expects a better 
determination of $\nu$ can be obtained by applying finite-size scaling analysis to $\rho_{s2} L$. 
Before presenting our results,
we would like to point out 
that since data from large volumes might be essential 
in order to determine the critical exponent $\nu$ accurately
as suggested in \cite{Jiang09.2}, we will use the strategy employed
in \cite{Jiang09.2} for our data analysis as well. 
A Taylor expansion of
Eq.~(\ref{FSS}) up to fourth order in $tL^{1/\nu}$ is used to fit the data of $\rho_{s2} L$. 
The critical exponent $\nu$ and critical point $(J'/J)_c$ calculated from the fit using
all the available data of $\rho_{s2} L $
are given by $0.6934(13)$ and $2.51962(4)$, respectively. The upper panel of 
figure \ref{fig2} 
demonstrates the result of the fit. Notice both $\nu$ 
and $(J'/J)_c$ we obtain are consistent with the corresponding results found in \cite{Wenzel08}.
By eliminating some data points of small $L$, we can reach a value of 
$0.700(3)$ for $\nu$ 
by fitting $\rho_{s2} L$ with $L \ge 26$ to Eq.~(\ref{FSS}). On the other 
hand, with the 
same range of $L$ ($L \ge 26$), a fit of $\rho_{s} L$ to Eq.~(\ref{FSS}) 
leads to $\nu = 0.688(2)$ and $(J'/J)_c = 2.5193(2)$,
both of which are consistent with those obtained in \cite{Wenzel08} as 
well (lower panel in
figure \ref{fig2}). By eliminating more data points of $\rho_{s} L $ with small $L$, 
the values for $\nu$ and $(J'/J)_c$ calculated from the fits are always consistent 
with those quoted above.
What we have shown clearly indicates that one would 
suffer the least correction by considering the finite-size scaling of the observable $\rho_{s2}L$.
As a result, it is likely
one can reach a value for $\nu$ consistent with the $O(3)$ prediction, namely 
$\nu=0.7112(5)$ if
large volume data points for $\rho_{s2}$ are available. Here we do not attempt to carry out
such task of obtaining data for $L > 64$. Instead, we employ the technique of fixing the 
ratio of spatial winding numbers squared in the simulations. 
Surprisingly, combining this new idea and finite-size scaling analysis, even from the observable $\rho_{s1} L$ which
is found to receive the most severe correction among the observables
considered here,
we reach a value for the critical exponent $\nu$ consistent with $\nu=0.7112(5)$ without 
additionally obtaining data points for $L > 64$. 
The motivation behind the idea of fixing the ratio of spatial winding numbers squared in the 
simulations is as follows. First of all, as we already mentioned earlier that the box size $L$ used in the simulations
is conventionally used as the $L^p$ in Eq. (\ref{FSS}) when carrying out finite-size scaling analysis. 
For isotropic systems, such strategy is no problem. However, for anisotropic cases, the validity of this
common wisdom of treating $L$ as $L^p$ is not clear. In particular, the same $L$ does not stand for the
same $L^p$ of the system for two different anisotropies $J'/J$.  Hence one needs to find a physical quantity 
which can really characterize the physical linear length of the system. 
Secondly, in magnon chiral perturbation theory which is the low-energy effective field theory for spin-1/2 antiferromagnets with $O(N)$ symmetry,
an exactly cubical space-time box is met when the condition $\beta c = L$ is satisfied, here
$c$ is the spin-wave velocity and $\beta$, $L$ are the inverse temperature and box size as before.
For spin-1/2 XY model on the square lattice, for large box size $L$,
the numerical value of c determined
by $L/\beta$ using the $\beta$ with which one obtains 
$\langle W^2 \rangle = 1/2(\langle W_1^2 \rangle + \langle W_2^2 \rangle) = \langle W_t^2 \rangle$ 
in the Monte Carlo simulations agrees quantitatively with the known results in the literature \cite{Jiang10.1}. 
This result implies that the squares of winding numbers are more physical than the box sizes since an exactly 
cubical space-time box is reached when the squares of spatial and temporal winding numbers are tuned to be the
same in the Monte Carlo simulations. Consequently the physical linear lengths of the system should be 
characterized by the squares
of winding numbers, not the  box sizes used in the simulations.
Based on what we have argued, it is
$\langle W^2_{1}\rangle/\langle W^2_{2}\rangle$, not $(L_2/L_1)^{2}$,
plays the role of the quantity $(L^p_2/L^p_1)^{2}$ for the system,
here again we refer $L^p_{i}$ with $i \in \{1,2\}$ as the physical linear length of the system in $i$-direction.
As a result,  fixing the ratio of spatial winding numbers squared in the simulations 
corresponds to the situation that the physical shape of the system remains fixed in all calculations.
Indeed it is demonstrated in \cite{Sandvik99} that rectangular lattice is more suitable 
than square lattice for studying the spatially anisotropic Heisenberg model with different
antiferromagnetic couplings $J_1$, $J_2$ in 1- and 2-directions. 
The idea of fixing the ratio of spatial winding numbers squared
quantifies the method used in \cite{Sandvik99}. 

\begin{figure}
\begin{center}
\includegraphics[width=0.36\textwidth]{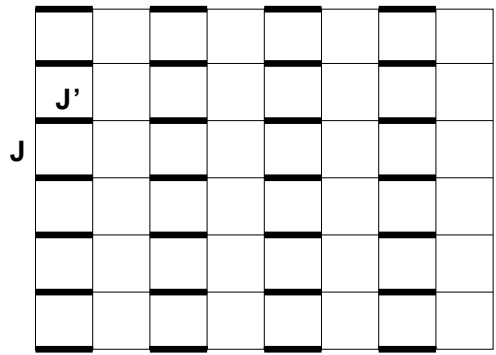}
\end{center}\vskip-0.5cm
\caption{Heisenberg model with a ladder pattern of anisotropy.}
\label{fig0.5}
\end{figure}

\begin{figure}
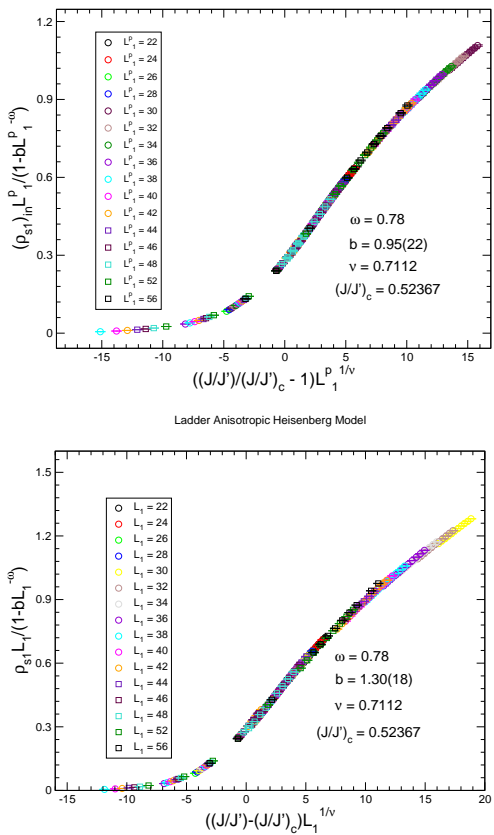

\begin{center}
\vbox{
\includegraphics[width=0.36\textwidth]{095_052367_07112_2.eps}\vskip0.25cm
\includegraphics[width=0.36\textwidth]{nr13.eps}
}
\end{center}\vskip-0.5cm
\caption{Comparison between the results of data collapse using the new unconventional 
finite-size scaling method (upper panel) described in the text and 
the conventional method (lower panel) for the ladder anisotropic Heisenberg model. 
The result in the upper panel is 
obtained from simulations with box sizes 
$(L_2-6) \times L_2,\, (L_2-4)\times L_2,\,...,\,(L_2+4) \times L_2$ for various
values of $J_2/J_1$ if the interpolations from such simulations
are reliable.}
\label{fig3}
\end{figure}

The method of fixing the ratio of 
spatial winding numbers squared is employed as follows. First of all, we 
perform a trial simulation to determine a fixed value for the ratio of spatial 
winding numbers squared which we denote by $w_f$ and will be used later 
in all calculations. Secondly, instead of fixing the aspect ratio of box sizes $L_1$ and $L_2$ in the 
simulations as in conventional finite-size scaling studies, we vary the 
variables $L_1$, $L_2$ and $J'/J$ in order to satisfy the condition of a 
fixed ratio of spatial winding numbers squared. This step involves a controlled
interpolation on the raw data points. In practice, for a fixed $L_2$ one performs simulations for a
sequence $L_1 = L_2, L_2\pm2, L_2\pm4, \dots$. The criterion of a fixed ratio of 
spatial winding numbers squared is reached by tuning the parameter $J_2/J_1$ 
and then carrying out a linear interpolation based on $(w/w_f)^{(-1/2)}$
for the desired observables, here
$w$ refers to the ratio of spatial winding numbers squared of the data 
points other than the trial one. Notice since only the ratio
of the physical linear lengths squared is fixed, 
it is natural to use $L_2$ in the 
finite-size scaling ansatz Eq.~(\ref{FSS}) both for the analysis of $\rho_{s1}$ 
and $\rho_{s2}$. 
The validity of this unconventional finite-size scaling method can be verified 
by considering the transition induced by dimerization for the Heisenberg model 
with a ladder pattern anisotropic couplings (figure \ref{fig0.5}). 
For $b \sim 0.95(22)$ in 
Eq.~(\ref{FSS}), we obtain a good data collapse for the observable 
$(\rho_{s1})_{\text{in}} L^p_1 (= (\rho_{s1})_{\text{in}}L_2)$. Above the 
subscript ``in'' means the data points are the interpolated one.
To make sure that the step of interpolation leads to accurate results,
we have carried out several trial simulations and have confirmed that
the interpolated data points are reliable as long as the ratio
is kept small (table 1). On the other hand, for $b=1.30(18)$ in 
Eq.~(\ref{FSS}),
a good data collapse is also obtained for the observable $\rho_{s1}L_1$, here
$\rho_{s1}$ are the raw data determined from the simulations directly.
Figure \ref{fig3} shows a comparison between the data collapse 
obtained by using the new unconventional method introduced above (upper panel) 
and by the conventional method (lower panel). For obtaining figure \ref{fig3}, 
we have fixed $\nu=0.7112$, $\omega = 0.78$, and $(J/J')_c = 0.52367$, which 
are the established values for these quantities. As one sees in figure 
\ref{fig3}, the quality of the data collapse obtained with the new method is 
better than the one obtained with the conventional method, thus confirming the 
validity of the idea to fix the ratio of winding numbers squared in order to 
studying the critical theory of a second order phase transition.

\begin{table}
\label{tab1}
\begin{center}
\begin{tabular}{ccccccc}
\hline
$J'/J$ & $L_1$  & $L_2$ & $w_f/w$ & $(\rho_{s1})_{{\text{in}}}$ & $\rho_{s1}$ \\
\hline
0.53 & 96 &  96 &    0.9558(33) &     0.008188(22) &    0.008198(7)\\
\hline
0.53 & 96 &  94 &    0.9549(32) &     0.008391(21) &    0.0084098(74)\\
\hline
0.545 & 90  & 94 &    0.9594(35) &     0.016862(33) &    0.016835(15)\\
\hline
0.545 & 90  & 90 &    0.9539(36) &     0.017651(35) &    0.017676(17)\\
\hline
0.535  & 98  & 98 &   0.9591(28)   &  0.011707(28)    &  0.0117297(124)  \\
\hline
0.54  &  96 & 96 &    0.9592(29) &    0.014838(37)  &   0.014846(13)  \\
\hline
0.525  & 96  & 96 &  0.9503(41)    & 0.0072255(225)    &  0.0072579(66)  \\
\hline
\end{tabular}
\end{center}
\caption{Comparison between interpolated and original values of $\rho_{s1}$ for several data
points. The data points which are used for interpolation are obtained from the simulations with 
$L_1 \times (L_2+2)$ (except the last row which is obtained from a simulation 
with $(L_1+2)\times L_2$). The inverse temperature $\beta$ for these data points are fixed to
$\beta J = 800$. }  
\end{table}

As demonstrated above, in general for a fixed $L_2$, one
can vary $L_1$ and $J'/J$ in order to reach the criterion of a fixed aspect-ratio of 
spatial winding numbers squared
in the simulations. For our study here,
without obtaining additional data, we proceed as follows. First of all, we calculate
the ratio 
$\langle W^2_{1}\rangle/\langle W^2_{2}\rangle$ for the data point at $J'/J = 2.5196$ and $L = 40$
which we denote by $w_f$.
We further choose $L^p_1 = L$ in our data analysis.
After obtaining this number,  
a linear interpolation for $\rho_{s1}$ of other data points based on $(w/w_{f})^{(-1/2)}$ 
is performed in order to reach the
criterion of a fixed ratio of spatial winding numbers squared in the 
simulations. The $w$ appearing above is again the corresponding 
$\langle W^2_{1}\rangle/\langle W^2_{2}\rangle$ 
of other data points. Here a controlled interpolation similar to what we have done in studying
the ladder anisotropic Heisenberg model is performed as well. 
Further, since large volumes data is 
essential for a quick convergence of $\nu$ as suggested in \cite{Jiang09.2},  
we make sure the set of interpolated data chosen for finite-size scaling analysis  
contains sufficiently many points
from large volumes as long as the interpolated results are reliable.
A fit of the interpolated $(\rho_{s1})_{\text{in}} L$ data to Eq.~(\ref{FSS}) with $\omega$ 
being fixed to its $O(3)$ value ($\omega = 0.78$) leads to
$\nu = 0.706(7)$ and $(J/J)_c = 2.5196(1)$ for $36 \le L \le 64$ (figure \ref{fig3}). 
Letting $\omega$ be a fit parameter results in consistent $\nu = 0.707(8)$ and 
$(J'/J)_c = 2.5196(7)$. Further, we always arrive at consistent results with 
$\nu = 0.706(7)$ and $(J'/J)_c = 2.5196(1)$ from the fits using $L > 36$ data. 
The value of $\nu$ we calculate from the fit is in good agreement 
with the expected $O(3)$ value
$\nu=0.7112(5)$. The critical point $(J'/J)_c = 2.5196(1)$ is consistent with that
found in \cite{Wenzel08} as well. To avoid any bias, we perform another analysis 
for the raw $\rho_{s1}L$ data with the same range of $L$ and $J'/J$ as we did for the interpolated data. 
By fitting 
this set of original data points to Eq.~(\ref{FSS}) with a fixed $\omega = 0.78$, 
we arrive at $\nu=0.688(7)$
and $(J'/J)_c = 2.5197(1)$ (figure \ref{fig4}), 
both of which again agree quantitatively with those determined in \cite{Wenzel08}. 
Similarly, applying this unconventional finite-size scaling to $\rho_{s2}$ would
lead to a numerical value of $\nu$ consistent with $\nu = 0.7112(5)$. For instance,
the $\nu$ determined by fitting $(\rho_{s2})_{\text{in}}L$ 
to Eq.~(\ref{FSS}) is found to be $\nu = 0.706(7)$, which agrees quantitatively with the 
predicted $O(3)$ value (figure \ref{fig6}).
Finally we would like to make a comment 
regarding the choice of $w_f$. In principle one can use $w_f$ determined from 
any $L$ and from any $J'/J$ close to $(J'/J)_c$. However it will be desirable 
to choose $w_f$ such that
the set of interpolated data used for analysis 
includes as many data points from large volumes as possible. 
Using the $w_f$ obtained at $J'/J = 2.5191$ ($J'/J = 2.5196$) with
$L = 40$ ($L = 44$), we reach the results of $\nu=0.704(7)$ and $(J'/J)_c = 2.5196(1)$ 
($\nu=0.705(7)$ and $(J'/J)_c = 2.5196(1)$) from the fit with a fixed
$\omega = 0.78$. These values for $\nu$ and $(J'/J)_c$ agree with what we have obtained earlier.
Indeed as we will demonstrate in another investigation, the critical exponent $\nu$
determined by the idea of fixing the ratio of spatial winding number squared in the simulations is
independence of the chosen reference point.


\section{Discussion and Conclusion}
\label{discussion}
\begin{figure}
\begin{center}
\includegraphics[width=0.365\textwidth]{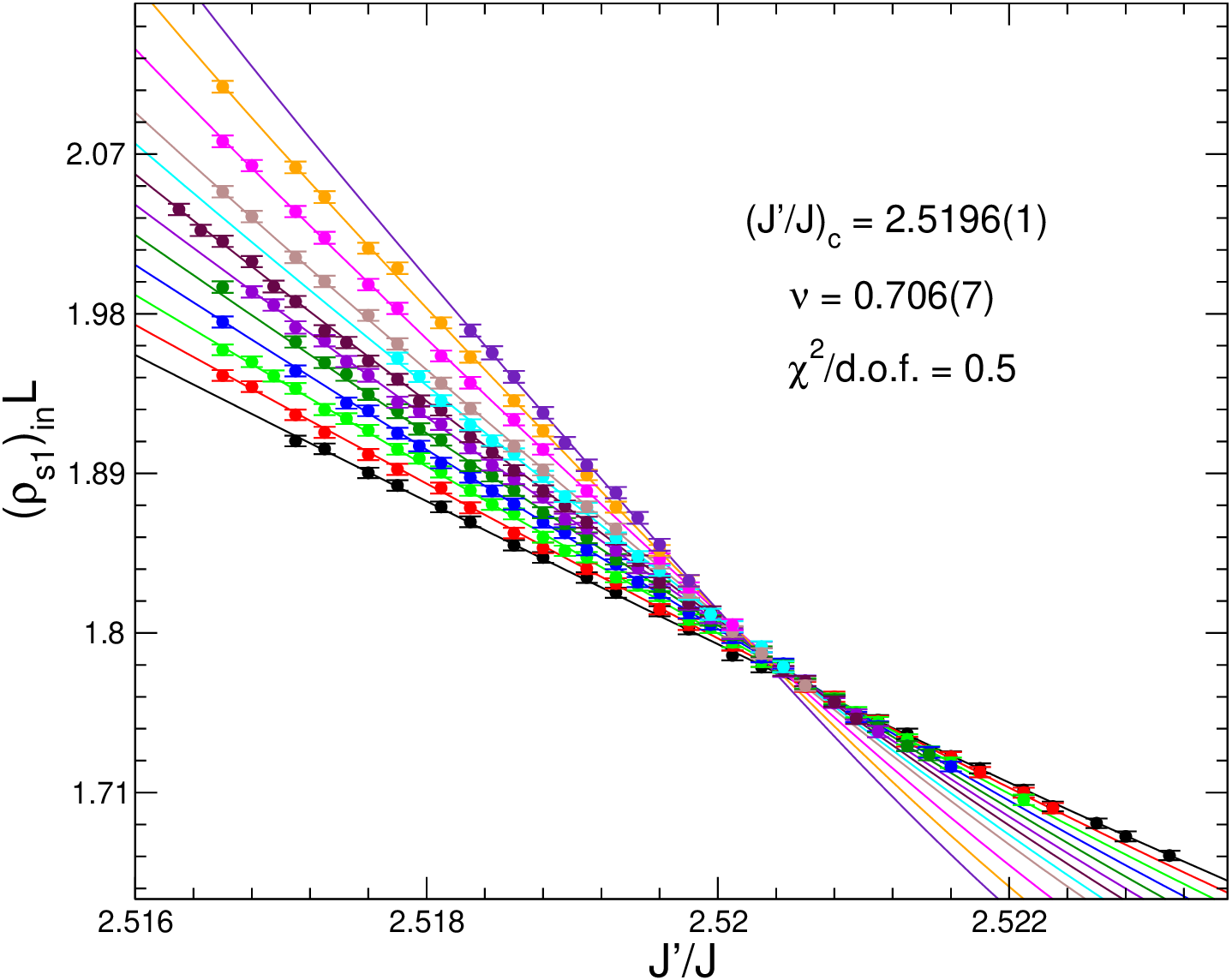}
\end{center}\vskip-0.5cm
\caption{Fit of interpolated $(\rho_{s1})_{{\text{in}}}L$ data
to Eq.~(\ref{FSS}). While the circles
are the numerical Monte Carlo data from the simulations, the solid curves 
are obtained by using the results from the fit.
}
\label{fig4}
\end{figure}

\begin{figure}
\begin{center}
\includegraphics[width=0.365\textwidth]{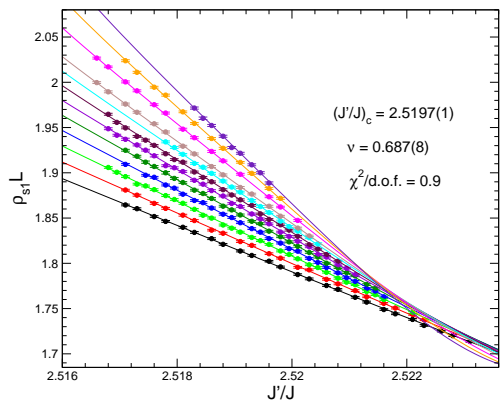}
\end{center}\vskip-0.5cm
\caption{Fit of original $\rho_{s1}L$ data
to Eq.~(\ref{FSS}). While the circles
are the numerical Monte Carlo data from the simulations, the solid curves
are obtained by using the results from the fit.
}
\label{fig5}\vskip-0.5cm
\end{figure}

\begin{figure}
\begin{center}
\includegraphics[width=0.365\textwidth]{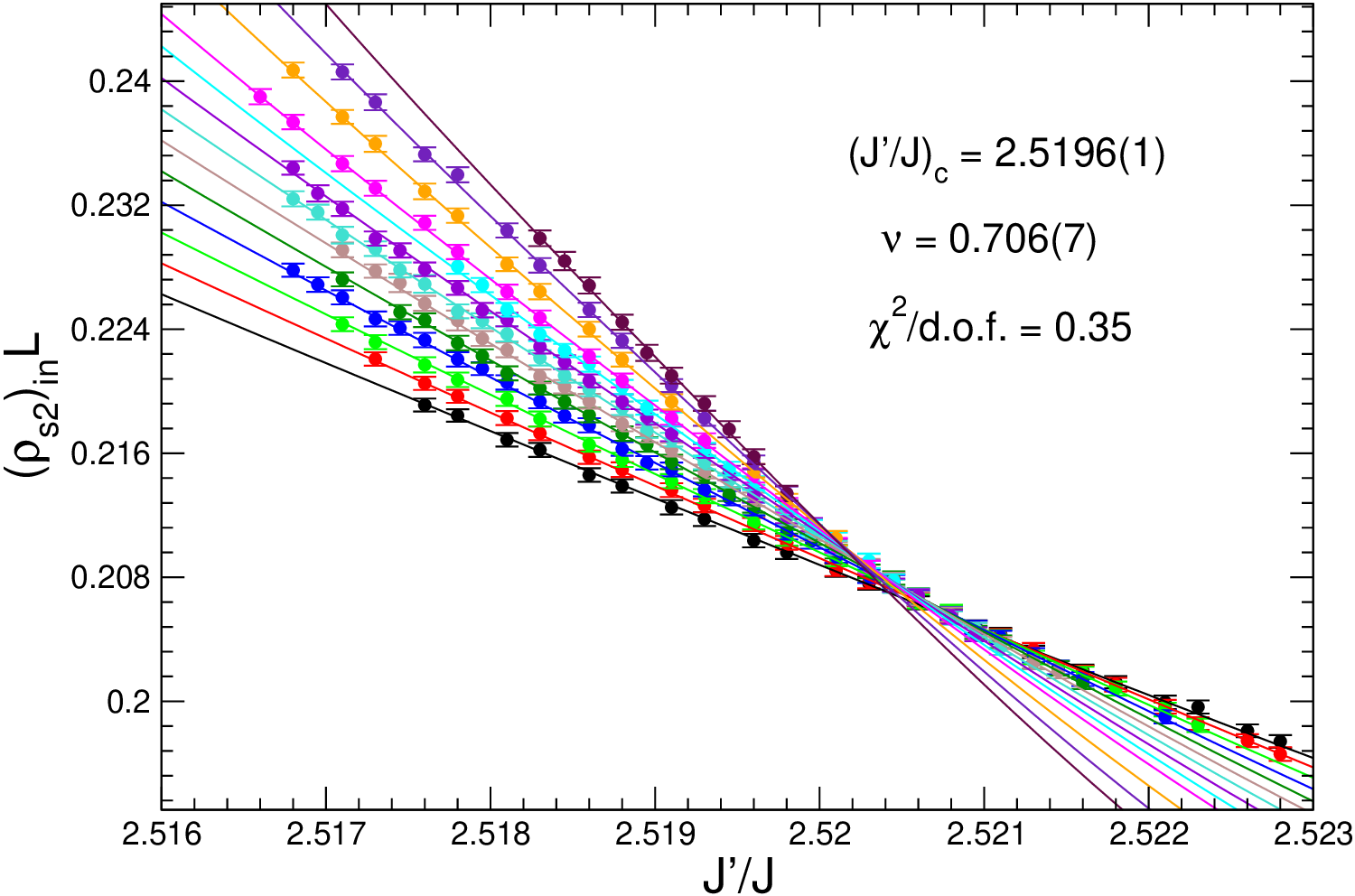}
\end{center}\vskip-0.5cm
\caption{Fit of interpolated $(\rho_{s2})_{{\text{in}}}L$ data
to Eq.~(\ref{FSS}). While the circles
are the numerical Monte Carlo data from the simulations, the solid curves 
are obtained by using the results from the fit.
}
\label{fig6}
\end{figure}

\vskip-0.2cm
In this paper, we revisit the phase transition driven by dimerization for the 
spin-1/2 Heisenberg model with
a spatially staggered anisotropy on the square lattice. 
We find that the observable $\rho_{s2} L$
suffers a much less severe correction compared to that 
of $\rho_{s1} L$, hence 
is a better quantity for finite-size scaling analysis. Further, 
we propose an unconventional finite-size scaling method,
namely we fix the ratio of spatial winding numbers squared. As 
a result, the physical shape of the system remains fixed in
all simulations and analysis. 
With this new strategy, we arrive at $\nu=0.706(7)$ for the critical exponent 
$\nu$ which is consistent with the most accurate Monte Carlo $O(3)$ result 
$\nu = 0.7112(5)$ by using only up to $L = 64$ data points derived from both $\rho_{s1} L$ and $\rho_{s2}L$.
Interestingly, the $\chi^2/{\text{d.o.f.}}$ obtained 
from the fits using the interpolated data are better than those resulted from the fits using 
the raw data (figures \ref{fig4}, \ref{fig5} and \ref{fig6}). This observation provides
another evidence to support the quantitative correctness of the new unconventional
finite-size scaling we proposed here.

It seems that when carrying out the finite-size scaling analysis for
the observables considered here, the use of physical linear lengths of the system, which are 
charaterized by the spatial winding numbers squared, would lead to a faster convergence of $\nu$.
It will be interesting to apply a similar technique to other observables such as Binder cumulants 
as well. However, for Binder cumulants, the correction from interpolation
will cancel out because of the definition of these observables. Therefore to further
test the philosophy behind the unconventional finite-size scaling method proposed here 
might require some new ideas. Nevertheless, with this new unconventional finite-size scaling method, 
we have 
successfully  
solved the puzzle raised in \cite{Wenzel08} by 
showing that the anisotropy driven phase transition for the spin-1/2 
Heisenberg model
with a staggered spatial anisotropy indeed belongs to the $O(3)$ universality 
class. Of course, the conventional finite-size scaling analysis is more convenient since
one does not need to carry out interpolation on the raw data. However for the 
subtle phase transition considered in this study, without obtaining data of
gigantic lattices, a new idea which is more physical oriented such as the one 
presented here is necessary. Still, to clarify the puzzle of an unconventional phase transition for the model studied here
as observed in \cite{Wenzel08} by simulating larger lattices and using the conventional finite-size scaling method is desirable. 
However, such investigation is beyond the scope of our study.

\section*{Acknowledgements}
\vskip-0.25cm
The simulations 
in this study are based on the loop algorithms available in ALPS 
library \cite{Troyer08} and were carried out
on personal desktops. Part of the results presented in this study 
has appeared in arXiv:0911.0653 and was  done at 
``Center for Theoretical Physics, Massachusetts Institute of Technology,
77 Massachusetts Ave, Cambridge, MA 02139, USA ``.  Partial support from DOE and NCTS (North)
as well as useful discussion with U.~J.~Wiese are acknowledged.


\end{document}